\documentclass[journal]{IEEEtran}
\usepackage[dvipsnames]{xcolor}
\pdfoutput=1
\usepackage{cite}
\usepackage{amsmath,amssymb,amsfonts}
\usepackage{algorithmic}
\usepackage{graphicx}
\usepackage{textcomp}
\usepackage[ruled,linesnumbered]{algorithm2e}
\usepackage{stmaryrd}
\usepackage{bm}

\def\BibTeX{{\rm B\kern-.05em{\sc i\kern-.025em b}\kern-.08em
    T\kern-.1667em\lower.7ex\hbox{E}\kern-.125emX}}

\newcommand{\argmi}[1]{\underset{#1}{\text{argmin}\ }}

\begin{document}
\title{One-step inversion algorithms for spectral CT, with application to dynamic Cone Beam CT.}
\author{Frédéric Jolivet, Georg Schramm, and Johan Nuyts

\thanks{This work is supported by the European Research Council (ERC) under the European Union’s Horizon 2020 Research and Innovations Programme (Grant Agreement No.780026).}
\thanks{F. Jolivet, G. Schramm and J. Nuyts  are with the Department of Imaging and Pathology, Division of Nuclear
Medicine, KU Leuven, Belgium. (e-mail: frederic.jolivet@kuleuven.be).}
}

\maketitle

\begin{abstract}
 Dual energy Cone Beam Comptuted Tomography (DE-CBCT) is a promising technique for several medical applications, including dynamic angiography. Recently, a dynamical two-step method has been proposed :  first, the water and iodine projections are computed from the multi-energy sinograms, then, a dynamic image of the iodine contrast is reconstructed using 4D Total-Variation (TV) constrained reconstruction from the iodine projections. In contrast to the 2-step methods, one-step methods use a model relating directly the multi-material images to the multi-energy sinograms. This kind of methods are well-known to reduce the noise correlation between the material images by avoiding the intermediate decomposition step, but request to solve an non-convex large scale optimization problem which can be challenging. In this work we use the Non-Linear Primal–Dual Hybrid Gradient Method (NL-PDHGM) optimization framework to propose two versions of a one-step method which is based on an empirical model : the first one is a version which considers the multi-material images as static object whereas the second version is developped for a spectific application with a static water image and a dynamic iodine image consisting of a series of 3D iodine images associated to different time points. This last version is developped to obtain the evolution of the iodine concentration in the blood vessels during a single CBCT scan. To evaluate the proposed one-step methods we used simulations which consider a CBCT system with dual layer spectral detector and a brain phantom with a static and a dynamic vascular tree. The proposed one-step methods are compared with 2-step methods.
\end{abstract}

\begin{IEEEkeywords}
 Inverse problems, Iterative Image reconstruction, One-step method, material decomposition, Spectral CT.
\end{IEEEkeywords}

\section{Introduction}
\label{sec:introduction}
Dual-energy CBCT data are of interest to medical applications, notably with the possibility to decompose the object onto some physical (photo-electric/compton,...) or materials basis (water/bone,water/iodine,...)~\cite{alvarez1976energy,zbijewski2014dual}. The material decomposition problem can be tackled by different strategies. First of all, in 2-step methods the materials projections are computed from the multi-energy sinograms, then, a reconstruction method (FDK, iterative methods,...) is used to reconstruct material specific images from the multi-material decomposed projections \cite{alvarez2011estimator}. In general, the material decomposition step greatly amplifies noise due to the ill-conditioning of the inversion step in the basis change. This makes the reconstruction in step 2 more challenging, because the material projections are corrupted by a high amount of correlated noise, which is hard to account for. To reduce this noise amplification effect several works proposed to introduce some constraints (non-negativity constraints, regularization...) in the projection domain \cite{ducros2017regularization,mechlem2018spectral,jolivet2021twostep}.  In contrast to the 2-step methods, one-step methods propose to solve the decomposition problem in a constrained one-step inversion, i.e. estimate multi-material reconstructions images from multi-energy sinograms  using a non-linear physical forward model \cite{cai2013full,long2014multi,nakada2015joint,zhao2014extended,barber2016algorithm,chen2017image,mory2018comparison} or based on an empirical forward model \cite{mechlem2017joint,jolivetIEEE2020}. An advantage of the one-step method is that it only needs to model the noise in the original sinograms, which is typically uncorrelated. Another benefit of this kind of methods in comparison to the 2-step methods, is that the one-step methods are able to perform a material decomposition from data acquired with Dual kV CT systems, with a potential mismatch between projections acquired with the high voltage source and projections acquired with the low voltage source \cite{rizzo2021reconstructing}. While the one-step method of \cite{jolivetIEEE2020} used a linear empirical model solving a convex optimization problem, most of the one-step methods use a non-linear forward model that can lead to solve a non-convex optimization problem which is a non-trivial challenge~\cite{chen2021non}. Some methods formulate a convex quadratic local bounding function to the non-convex data discrepancy term and use a convex primal dual optimization algorithm to solve the local quadratic approximation \cite{barber2016algorithm,barber2016mocca}, whereas some methods used a convexification of the data fidelity term \cite{barber2020convergence,chen2017image,chen2018algorithm}. In this paper we present in Sec.\ref{Sec:staticonestep} a one-step method using an empirical polynomial model of order 2 which leads to a possibly non-convex optimization problem which is solved by the Non-Linear Primal–Dual Hybrid Gradient Method (NL-PDHGM)~\cite{valkonen2014primal} which is a non-linear adaptation of the Chambolle-Pock method \cite{chambolle2011first}. In \cite{valkonen2014primal} the author gives a local convergence proof of the method, provided various technical conditions are satisfied. This proposed one-step method applies non-negativity and sparsity constraints including a Total-Variation (TV) regularization.  \\
One the other hand, for standard CT applications several works previously published have proposed approaches for dynamic reconstructions based on the 4D TV regularization with different medical applications in cardiac, thoracic, pulmonary and brain imaging \cite{hansen2018fast,mory2016motion,ritschl2012iterative,li2018time}. These methods require to solve a non-smooth large-scale optimization problem, therefore it is crucial to use an efficient optimization strategy to have an acceptable computation time. In the last decade, many works proposed computationally efficient implementations based on the primal-dual optimization algorithm of Chambolle and Pock \cite{chambolle2011first} for dynamic reconstructions \cite{mory2014modified,taubmann2017assessing,nikitin2019four}. Recently we have proposed a dynamical iodine reconstruction based on a 2-step method with data acquired with dual-energy (DE) CBCT devices \cite{heylen20194d} with a motion-correction extension~\cite{heylenmotion}. 

Although the aforementioned one-step methods consider the object as static during the scan, we present in Sec.\ref{Sec:dynamiconestep} an extension of the proposed one-step method for dynamic reconstruction. The proposed dynamical one-step method considers the water image as a static object and the iodine image as a 4D image, represented by a temporal sequence of 3D images. As in the proposed static one-step method, the proposed dynamical one-step method uses an empirical polynomial model of order 2 and the NL-PDHGM optimization framework proposed by Valkonen \cite{valkonen2014primal}. Our objective is to create a dynamic iodine reconstruction from a single spectral CBCT scan, which can be used to visualize the flow of contrast agent through the brain vasculature, which has a large diagnostic potential in the acute ischemic stroke workflow. 

To distinguish the two versions of the one-step method proposed in this manuscript, the version which considers all material images as static will be named ”static one-step method” whereas the extended version which considers the water image as static and the iodine image as a 4D image will be named ”dynamic one-step method”.   

 To assess the capabilities of these proposed one-step methods, we simulate data of a dual-energy angiographic CBCT-scan of a brain phantom, where the iodine concentration in the blood vessels is fixed for simulation 1 detailed in Sec.\ref{sec:simustatic}. Simulation 2, detailed in Sec.\ref{sec:simudynamic}, is obtained with a dynamic brain phantom where the iodine concentration changes during the CBCT scan. For all these simulations we consider a CBCT system that obtains dual-energy data by using a stack of two detector layers, where the first layer acts as an energy dependent filter for the second. Section \ref{sec:Resultstatic} presents the static one-step method reconstructions (from simulation 1 data), comparing them to those of  a few different 2-step methods. In section \ref{sec:Resultdynamic}, the reconstructions obtained with the dynamic one-step method are compared to those of the dynamical 2-step method of~\cite{heylen20194d}.
 The paper will finish with some discussions in Sec.\ref{sec:discussion}; we conclude and give some perspectives in Sec.\ref{sec:conclusion}.

\section{Methods}
\subsection{An empirical forward model}
The continuous theorical model for dual energy CT data can be expressed as,
\begin{equation}
{m}_c(l_{w},l_{i})=-\log\left(\dfrac{\int_E W_c(E) \exp{(-\mu_i(E) l_{i} -\mu_w(E) l_{w})   } }{\int_E W_c(E)}\right) 
\label{eq:modelth}
\end{equation}
where $c$ is the index of the detector layer, $\mu_{i}(.)$ and $\mu_{w}(.)$ are the material basis functions associated with iodine and water, and $l_i$ and $l_w$ are the corresponding equivalent thicknesses. The function $W_c(.)$ is expressed as $W_c(E)=E~\phi(E)S_c(E)$
with $E$ the energy, $\phi(.)$ is the energy source spectrum and $S_c(.)$ is the function of the detector spectral sensitivity.\\
While a discrete version of \eqref{eq:modelth} can be used, in this work we consider an empirical model which estimates the expectation of the log-converted measured dual-energy sinogram as follows,
\begin{equation}
\Tilde{m}_c(l_{w},l_{i})=a_{5c} l_w^2+ a_{4c}l_i^2 + a_{3c}l_wl_i + a_{2c}l_w + a_{1c}l_i
\label{eq:model}
\end{equation}
where the polynomial coefficients $a_c$ are estimated by fitting a set of attenuation values observed by each detector layer, for different combinations of water and iodine thicknesses. These attenuation values can be obtained with calibrated data \cite{alvarez2011estimator,mechlem2017joint,jolivetIEEE2020} or calculated using a physical model which requires knowledge of the source spectrum and the detector response. In this work, where the method is evaluated with simulations, we use the latter strategy. For dual-energy CBCT data, we define $\Tilde{\textbf{m}}_c$ the vectorized version of the empirical model \eqref{eq:model}. Therefore, each energy layer sinogram $\textbf{s}_c \in \mathbb{R}^M$ can be expressed as,
\begin{equation}
\textbf{s}_c=\Tilde{\textbf{m}}_c\left(\textbf{l}_w,\textbf{l}_i\right)+\textbf{e}_c
\label{eq:vecmodel}
\end{equation}
where $\textbf{l}_i \in \mathbb{R}^M$ (respectively $\textbf{l}_w \in \mathbb{R}^M$) are the iodine projections (respectively the water projections) and $\textbf{e}_c \in \mathbb{R}^M$ is the error vector between measurements and the empirical model (including detection noise, electronic noise and modeling errors).

\subsection{A static one-step inversion} \label{Sec:staticonestep}

Most of the time, one-step methods consider a static object~\cite{mory2018comparison}. In a static one-step method we consider that the iodine projections and the water projections are defined as $\textbf{l}_i=\textbf{A}\textbf{x}_i$ and $\textbf{l}_w=\textbf{A}\textbf{x}_w$ where $\textbf{x}_i \in \mathbb{R}^N$ and $\textbf{x}_w \in \mathbb{R}^N$ are the iodine image and water image, whereas $\textbf{A}\in \mathbb{R}^{M\times N}$ denotes the forward tomographic projector matrix. Then, we define a vector of unknown elements $\textbf{x} \in \mathbb{R}^{2N}$ and a continuous non-linear operator $K_0 : \mathbb{R}^{2N} \rightarrow \mathbb{R}^{2M}$ such that,
\begin{equation}
    \textbf{x}=\begin{pmatrix} \textbf{x}_w\\ \textbf{x}_i
    \end{pmatrix} \quad \text{and} \quad
    K_0(\textbf{x})=\begin{pmatrix}  \Tilde{\textbf{m}}_{1}\left(\textbf{A}\textbf{x}_w,\textbf{A}\textbf{x}_i\right) \\
\Tilde{\textbf{m}}_{2}\left(\textbf{A}\textbf{x}_w,\textbf{A}\textbf{x}_i\right)
    \end{pmatrix}.
\end{equation}

Therefore, assuming error vectors $\textbf{e}_{c=1,2}$ as non-correlated Gaussian noise with constant variance in \eqref{eq:vecmodel}, the data fidelity term of the conventional one-step approach (proportional to the negative log-likelihood), will be expressed~as,
\begin{equation}
\textit{F}_0\left(K_0(\textbf{x})\right) = ~\left\Vert K_0(\textbf{x})-\textbf{s}\right\Vert_2^2
\label{eq:data_fidelity}
\end{equation}
with $\textit{F}_0(\textbf{y}_0) =\Vert \textbf{y}_0 -\textbf{s}\Vert_2^2 $ where $\textbf{s} \in \mathbb{R}^{2M}$ are tomographic dual-energy data such that $\textbf{s}=\begin{pmatrix}
    \textbf{s}_1\\
    \textbf{s}_2
\end{pmatrix} $.

One of the key points of one-step methods is to introduce prior information (non-negativity constraints, regularizations...) in the image domain to constrain the method and lead it to a satisfying solution. For this static one-step method, we consider isotropic total-variation constraints on the water and the iodine which are termed as two regularization functions $F_1(K_1(\textbf{x}))$ and $F_2(K_2(\textbf{x}))$. Therefore we introduce two continuous linear operators $K_1 : \mathbb{R}^{2N} \rightarrow \mathbb{R}^{3N}$ and $K_2 : \mathbb{R}^{2N} \rightarrow \mathbb{R}^{3N}$ such that $K_1(\textbf{x})=\nabla_{3D} \textbf{x}_w$ and  $K_2(\textbf{x})=\nabla_{3D} \textbf{x}_i$ where $\nabla_{3D}$ represents the 3D finite difference. For example, if we transform a vector $\bm{\nu} \in \mathbb{R}^{N}$ as a 3D image $\bm{\nu}_{j,k,l}$, then the operator $\nabla_{3D}$ applied on $\bm{\nu}$ gives 3 components $\mathbf{u}_1$, $\mathbf{u}_2$ and $\mathbf{u}_3$ such that,
\begin{equation}
\left(\nabla_{3D}~\bm{\nu}\right)_{j,k,l}= \left\{ \begin{array}{lll}
     \textbf{u}_{1,j,k,l}=\bm{\nu}_{j,k,l}-\bm{\nu}_{j-1,k,l}  \\
     \textbf{u}_{2,j,k,l}=\bm{\nu}_{j,k,l}-\bm{\nu}_{j,k-1,l} \\
     \textbf{u}_{3,j,k,l}=\bm{\nu}_{j,k,l}-\bm{\nu}_{j,k,l-1} . 
\end{array}\right.
\label{eq:diffin}
\end{equation}
On the other hand, $F_1$ and $F_2$ are functions defined as
\begin{equation}
 F_1(\textbf{y}_1)=\alpha_1\Vert \textbf{y}_1\Vert_{2,1} \quad \text{and} \quad F_2(\textbf{y}_2)=\alpha_2\Vert \textbf{y}_2\Vert_{2,1}
\label{eq:mixednorm}
\end{equation}
where  $\alpha_1 \in \mathbb{R}$ and $\alpha_2 \in \mathbb{R}$ are regularization hyper-parameters, and $\Vert .\Vert_{2,1}$ represents the mixed (2,1)-norm which can be expressed following the notation in \eqref{eq:diffin},
\begin{equation}
\Vert \textbf{u}\Vert_{2,1}=\sum_{j,k,l} \sqrt{\sum_d \textbf{u}_{d,j,k,l}^2}  .
\label{eq:mixednormdef}
\end{equation}
For the static one-step method we want to design a regularization for our specific static application, i.e to be able to see the iodine concentration in the blood vessels from a dual energy CBCT data. That is why we include a sparsity constraint on the iodine image which promotes blood vessels which have sparse structures, whereas we include a non-negativity constraint on iodine and water images to help the material decomposition and limit the anti-correlated noise between materials. Therefore we define two functions $G_1$ and $G_2$ expressed as $G_1(\textbf{x})=\mathcal{X}_{\geq \mathbf{0}}\left(\textbf{x}_w\right)$ and $ G_2(\textbf{x})=\alpha_3 \Vert \textbf{x}_i\Vert_1 + \mathcal{X}_{\geq \mathbf{0}}\left(\textbf{x}_i\right)$ where $\alpha_3 \in \mathbb{R}$ is a regularization hyper-parameter, and $\mathcal{X}_{\geq \mathbf{0}}$ is the indicator function defined as, 
\begin{equation}
\mathcal{X}_{\geq \mathbf{0}}\left(\bm{\nu}\right)_{j,k,l} = \left\{ \begin{array}{ll}
& 0 \hspace{8mm} \textbf{if} \quad \bm{\nu}_{j,k,l} \geq 0 \vspace{0.5mm} \\
& +\infty \quad \textbf{if} \quad \bm{\nu}_{j,k,l} < 0.
\end{array} \right.
\label{eq:indicatorpositivity}
\end{equation}
Including all these constraints, the proposed static one-step method can be expressed as the following optimization problem,
\begin{equation}
  \hat{\textbf{x}} \in  \argmi{\textbf{x}\in \mathbb{R}^{2N}}~ \sum_{h=0}^2 F_h\left(K_h(\textbf{x})\right) + \sum_{\xi=1}^2 G_\xi(\textbf{x})
  \label{eq:staticoptimpb}
\end{equation}
In Valkonen's optimization framework \cite{valkonen2014primal}, the primal-dual
formulation of the nonlinear primal problem \eqref{eq:staticoptimpb} can be expressed as,
\begin{equation}
  \hat{\textbf{x}} \in  \argmi{\textbf{x}} \sum_{h=0}^2~ \underset{\textbf{y}_h}{\text{max}} \langle  K_h(\textbf{x}),\textbf{y}_h\rangle - \textit{F}^*_h\left(\textbf{y}_h\right) + \sum_{\xi=1}^2 G_\xi(\textbf{x})
  \label{eq:minmaxpb}
\end{equation}
where $\textbf{y}_0 \in \mathbb{R}^{2M}$, $\textbf{y}_1 \in \mathbb{R}^{3N}$, $\textbf{y}_2 \in \mathbb{R}^{3N}$ are the dual variables and for all h=\{0,1,2\} the function $\textit{F}_h^*$ is the convex conjugate of the function $\textit{F}_h$.\\
To find a saddle point of the primal-dual optimization problem~\eqref{eq:minmaxpb} we use the Exact NL-PDHGM framework \cite{valkonen2014primal}. Because \eqref{eq:staticoptimpb} is an optimization problem which can be non-convex, the Exact NL-PDHGM framework only guarantees convergence to a local minimum, which may differ from the global one. Algorithm \ref{algostatic}  presents the proposed static one-step method which solves the reconstruction problem \eqref{eq:staticoptimpb} using the Exact NL-PDHGM framework of \cite{valkonen2014primal}. Algorithm \ref{algostatic} was obtained by inserting our problem in equations (2.4a) - (2.4c) of \cite{valkonen2014primal} as follows~:\\
$\bullet$ Eq. $y^{n+1}=\left(I +\sigma \partial F^*\right)^{-1}\left(y^n+ \sigma K(x^n_\omega) \right)$ (2.4c) :
\begin{equation*}
    \left|
    \begin{array}{ll}
       \textbf{y}^{n+1}_0=\left(I +\sigma_0 \partial F_0^*\right)^{-1}\left(\textbf{y}^n_0+ \sigma_0 K_0(\bar{\textbf{x}}^n) \right) \quad \text{(Alg.\ref{algostatic}-l.\ref{grad0static1}-\ref{grad0static2})}  \\ \\
       \textbf{y}^{n+1}_1=\left(I +\sigma_1 \partial F_1^*\right)^{-1}\left(\textbf{y}^n_1+ \sigma_1 K_1(\bar{\textbf{x}}^n) \right)  \quad \text{(Alg.\ref{algostatic}-l.\ref{projball1})}\\ \\
       \textbf{y}^{n+1}_2=\left(I +\sigma_2 \partial F_2^*\right)^{-1}\left(\textbf{y}^n_2+ \sigma_2 K_2(\bar{\textbf{x}}^n) \right)  \quad \text{(Alg.\ref{algostatic}-l.\ref{projball2})}
    \end{array}\right.
\end{equation*}
$\bullet$ Eq. $x^{n+1}=\left(I +\tau \partial G \right)^{-1}\left(x^n- \tau [\nabla K(x^n_\omega)]^*y^{n+1} \right)$ (2.4a)~:
\begin{equation*}
  \left|
    \begin{array}{ll}
    \textbf{x}^{n+1}_w=\left(I +\tau \partial G_1 \right)^{-1}\left(\textbf{x}^n_w- \tau (\textbf{v}_{0,w}^{n+1}+\textbf{v}_1^{n+1})\right)~\text{(Alg.\ref{algostatic}-\ref{proj+})}\\
    \textbf{x}^{n+1}_i=\left(I +\tau \partial G_2 \right)^{-1}\left(\textbf{x}^n_i- \tau (\textbf{v}_{0,i}^{n+1}+\textbf{v}_2^{n+1})\right)~\text{(Alg.\ref{algostatic}-l.\ref{projsoft})}
    \end{array}\right.
\end{equation*}
where $\textbf{v}_1^{n+1}=K_1^*\textbf{y}^{n+1}_1$, $\textbf{v}_2^{n+1}=K_2^*\textbf{y}^{n+1}_2$ and $\textbf{v}_0^{n+1}=[\nabla K_0(\bar{\textbf{x}}^n)]^*\textbf{y}_0^{n+1}$ with $K_1^*$, $K_2^*$ and $[\nabla K_0(\bar{\textbf{x}}^n)]^*$ the adjoint operators of $K_1$, $K_2$ and $[\nabla K_0(\bar{\textbf{x}}^n)]$. While expressions of $\textbf{v}_1^{n+1}$ and $\textbf{v}_2^{n+1}$ are straightforward because $K_1$ and $K_2$ are linear operators and the adjoint of finite difference  operators is well-know as the negative divergence (see Alg.\ref{algostatic}-l.\ref{div1}-\ref{div2}), the expression of $\textbf{v}_0^{n+1}$ (Alg.\ref{algostatic}-l.\ref{grad1static}-\ref{grad2static}) is more complicate due to the non-linearity of the operator $K_0$. In the Appendix we present a proof of the expression of $\textbf{v}_0^{n+1}$.\\
$\bullet$ Eq. $x^{n+1}_\omega=x^{n+1}+\omega (x^{n+1}-x^n)$ (2.4b)  :
\begin{equation*}
\left|
    \begin{array}{ll}
     \bar{\textbf{x}}^{n+1}_w = \textbf{x}_w^{n+1} + \omega \left( \textbf{x}_w^{n+1} - \textbf{x}_w^{n}\right) \quad \text{(Alg.\ref{algostatic}-l.\ref{omeg1})}\\
     \bar{\textbf{x}}^{n+1}_i = \textbf{x}_i^{n+1} + \omega \left( \textbf{x}_i^{n+1} - \textbf{x}_i^n\right) \quad \text{(Alg.\ref{algostatic}-l.\ref{omeg2})}
    \end{array}\right.
\end{equation*}

\begin{algorithm}\label{algostatic}
\caption{The static one-step algorithm}
\SetAlgoLined

 Initialize all variables, choose $\omega \in [0, 1]$ and $\tau$, $\sigma_h\geq 0$ such that $\tau \sigma_h \Vert K_h\Vert^2 <1$ \;
 \For{n = 0 to niter-1}{
     $\textbf{y}_{0,1}^{n+1}=\dfrac{2}{2+\sigma_0}\left( \textbf{y}_{0,1}^{n}+\sigma_0 \left(\Tilde{\textbf{m}}_1\left(\textbf{A}\bar{\textbf{x}}_w^n,\textbf{A}\bar{\textbf{x}}_{i}^n\right)- \textbf{s}_{1} \right)\right)$ \label{grad0static1}\\
     $\textbf{y}_{0,2}^{n+1}=\dfrac{2}{2+\sigma_0}\left( \textbf{y}_{0,2}^{n}+\sigma_0 \left(\Tilde{\textbf{m}}_2\left(\textbf{A}\bar{\textbf{x}}_w^n,\textbf{A}\bar{\textbf{x}}_{i}^n\right)- \textbf{s}_{2} \right)\right)$ \label{grad0static2}\\    
     $\textbf{y}_{1}^{n+1}=proj_{\alpha_1 P} \left(\textbf{y}_{1}^{n} +\sigma_1 \nabla_{3D}\bar{\textbf{x}}_w^n \right)$ \label{projball1}\\
     $\textbf{y}_{2}^{n+1}=proj_{\alpha_2 P} \left(\textbf{y}_{2}^{n} +\sigma_2 \nabla_{3D}\bar{\textbf{x}}_{i}^n \label{projball2}\right)$\\
      $\textbf{v}_{0,w}^{n+1} =\textbf{A}^\mathsf{T}(\underset{c=1}{\overset{2}{\sum}}(2a_{5c}\textbf{A}\bar{\textbf{x}}_w^n+a_{3c}\textbf{A}\bar{\textbf{x}}_{i}^n+a_{2c}\textbf{1})\odot \textbf{y}_{0,c}^{n+1})$ \label{grad1static}\\
      $\textbf{v}_{0,i}^{n+1} = \textbf{A}^\mathsf{T}(\underset{c=1}{\overset{2}{\sum}}(2a_{4c}\textbf{A}\bar{\textbf{x}}_{i}^n+a_{3c}\textbf{A}\bar{\textbf{x}}_w^n+a_{1c}\textbf{1})\odot \textbf{y}_{0,c}^{n+1})$\label{grad2static}\\
     $\textbf{v}_{1}^{n+1}=-\operatorname{div}\left( \textbf{y}_{1}^{n+1}\right)$\label{div1}\\
     $\textbf{v}_{2}^{n+1}=-\operatorname{div}\left(\textbf{y}_{2}^{n+1}\right)$\label{div2}\\
     $\textbf{x}_w^{n+1}=proj_{\mathbb{R}^{N+}}\left(\textbf{x}_w^n-\tau \textbf{v}_{0,w}^{n+1}-\tau \textbf{v}_{1}^{n+1}\right)$\label{proj+}\\
     $\textbf{x}_{i}^{n+1}=\textit{S}^+_{\tau\alpha_3}\left(\textbf{x}_{i}^n-\tau \textbf{v}_{0,i}^{n+1}-\tau \textbf{v}_{2}^{n+1}\right)$ \label{projsoft} \\
     $\bar{\textbf{x}}^{n+1}_w = \textbf{x}_w^{n+1} + \omega \left( \textbf{x}_w^{n+1} - \textbf{x}_w^{n}\right)$ \label{omeg1}\\
     $\bar{\textbf{x}}^{n+1}_i = \textbf{x}_i^{n+1} + \omega \left( \textbf{x}_i^{n+1} - \textbf{x}_i^n\right)$ \label{omeg2}
    }
\end{algorithm} 
The operator $\odot$ in lines \ref{grad1static}-\ref{grad2static} represents the element-wise product (also known as the Hadamard product). In lines \ref{projball1}-\ref{projball2} the projection on the set $proj_{\alpha_i P}$ projects each voxel-wise onto the $\ell_2$-ball of radius $\alpha_i$, while in line \ref{projsoft} the positive soft-thresholding operator $\textit{S}^+_{\alpha_3}$ is applied voxel-wise~:
\begin{equation}
\textit{S}^+_{\alpha_3}\left(\bm{\nu}\right)_{j,k,l} = \left\{ \begin{array}{ll}
&\bm{\nu}_{j,k,l}-\dfrac{\alpha_3}{2} \quad \textbf{if} \quad \bm{\nu}_{j,k,l}> \dfrac{\alpha_3}{2} \vspace{0.5mm} \\
&0 \hspace{18.5mm} \textbf{if} \quad \bm{\nu}_{j,k,l}\leq \dfrac{\alpha_3}{2}.
\end{array} \right.
\label{eq:soft}
\end{equation}
In line \ref{proj+}, $proj_{\mathbb{R}^{N+}}$ enforces each element of a vector in $\mathbb{R}^N$ to be positive.

\subsection{A dynamic one-step inversion} \label{Sec:dynamiconestep}
In this section we propose an extension of the one-step method proposed in Sec.\ref{Sec:staticonestep} with a dynamical one-step method, which considers the water material as static during the entire scan, whereas the iodine is considered as dynamic. That is why in this dynamic one-step method the water projections and the iodine projections are respectively defined as $\textbf{l}_w=\textbf{A}\textbf{x}_w$ and $\textbf{l}_i=\Tilde{\textbf{A}}\textbf{x}_{i,.}$ where $\textbf{x}_{i,.} \in \mathbb{R}^{NT}$ is the 4D iodine image. This image is represented with a set of 3D volumes, one for each time point $t = t_1...t_T$, where we assume that the image for a particular time can be computed with linear interpolation between the two volumes at the two closest time points. $\Tilde{\textbf{A}} \in \mathbb{R}^{M\times NT}$ denotes the dynamical forward tomographic projector matrix which can be expressed as,
\begin{equation}
    \Tilde{\textbf{A}}=\begin{pmatrix}
    \textbf{A}_1 \textbf{Q}_1\\
    \textbf{A}_2 \textbf{Q}_2\\
    \vdots \\
    \textbf{A}_P \textbf{Q}_P\\
    \end{pmatrix} \quad \text{and the data } \quad \textbf{s}_c=\begin{pmatrix}
    \textbf{s}_{c,1}\\
    \textbf{s}_{c,2}\\
    \vdots \\
    \textbf{s}_{c,P}\\
    \end{pmatrix} ,
\end{equation}
where $P$ is the total number of tomographic projections, $\textbf{s}_{c,p} \in \mathbb{R}^{\frac{M}{P}}$ is the projection with index $p$ measured by the detector layer $c$, $\textbf{A}_p \in \mathbb{R}^{\frac{M}{P}\times N}$ denotes the forward tomographic projector matrix associated to the projection index $p$ whereas $\textbf{Q}_p\in~\mathbb{R}^{N\times NT}$ is a linear interpolator along the time dimension associated to the projection index $p$ \cite{mory2014modified,mory2016motion,jolivet2021dynamic}. For example, if $\textbf{x}_{i,.}$ contains ten time frames ($T=10$) and data of the $p-th$ projection $\textbf{s}_{c,p}$ has been acquired at the phase $\frac{p-1}{P-1}=0.47$, then $\textbf{Q}_{p}\textbf{x}_{i,.}=0.3 \textbf{x}_{i,4}+0.7 \textbf{x}_{i,5}$.\\
Then, we define a vector of unknown elements $\Tilde{\textbf{x}} \in \mathbb{R}^{(T+1)N}$ and a continuous non-linear operator $\Tilde{K}_0 : \mathbb{R}^{(T+1)N} \rightarrow \mathbb{R}^{2M}$ such that,
\begin{equation}
    \Tilde{\textbf{x}}=\begin{pmatrix} \textbf{x}_w\\ \textbf{x}_{i,.}
    \end{pmatrix} \quad \text{and} \quad
    \Tilde{K}_0(\Tilde{\textbf{x}})=\begin{pmatrix}  \Tilde{\textbf{m}}_{1}(\textbf{A}\textbf{x}_w,\Tilde{\textbf{A}}\textbf{x}_{i,.}) \\
\Tilde{\textbf{m}}_{2}(\textbf{A}\textbf{x}_w,\Tilde{\textbf{A}}\textbf{x}_{i,.})
    \end{pmatrix}.
\end{equation}
The data fidelity of the proposed dynamical one-step method is defined as,
\begin{equation}
\Tilde{\textit{F}}_0(\Tilde{K}_0(\tilde{\textbf{x}})) = \Vert\Tilde{K}_0(\tilde{\textbf{x}})-\textbf{s}\Vert_2^2
\label{eq:dyn_data_fidelity}
\end{equation}
with $\Tilde{\textit{F}}_0(\tilde{\textbf{y}}_0) =\Vert \tilde{\textbf{y}}_0 -\textbf{s}\Vert_2^2 $ where $\textbf{s} \in \mathbb{R}^{2M}$ are tomographic dual-energy data defined above \eqref{eq:data_fidelity}.\\
In this work, we aim to reconstruct the dynamic iodine image and the static water image from a single CBCT acquisition over 200 degrees. In our example with ten time frames (T=10) and a CBCT acquisition over 200 degrees, each time frame $\textbf{x}_{i,t}$ is linked only with projections over $\frac{200}{T}=20$ degrees. This problem is severely ill posed, so good spatio and temporal regularization is mandatory. That is why for this dynamical one-step we consider a 3D isotropic total-variation constraint on the 3D water image which is termed as the regularization function $\Tilde{F}_1(\Tilde{K}_1(\textbf{x}))$ and a 4D isotropic total variation constraint on the 4D iodine image which is termed as the regularization function $\Tilde{F}_2(\Tilde{K}_2(\textbf{x}))$. This 4D isotropic total-variation gives  a different  weight  in  the  time  direction. Therefore, we introduce two continuous linear operators $\Tilde{K}_1 : \mathbb{R}^{(T+1)N} \rightarrow \mathbb{R}^{3N}$ and $K_2 : \mathbb{R}^{(T+1)N} \rightarrow \mathbb{R}^{4TN}$ such that $\Tilde{K}_1(\tilde{\textbf{x}})=\nabla_{3D} \textbf{x}_w$ and $\Tilde{K}_2(\tilde{\textbf{x}})=\nabla_{4D}^\gamma \textbf{x}_{i,.}$ where $\nabla_{3D}$ represents the conventional finite difference operator for a 3D volume defined in \eqref{eq:diffin}, whereas $\nabla_{4D}^\gamma$ applies  a  finite  difference  in  four  dimensions, but  with  a different weight for the time dimension. For example, if we transform a vector $\bm{\nu} \in \mathbb{R}^{TN}$ as a 4D image $\bm{\nu}_{j,k,l,t}$, then the operator $\nabla_{4D}^\gamma$ applied on $\bm{\nu}$ gives 4 components $\mathbf{u}_1$, $\mathbf{u}_2$, $\mathbf{u}_3$, and $\mathbf{u}_4$ such that,
\begin{equation*}
\left(\nabla_{4D}^\gamma~\bm{\nu}\right)_{j,k,l,t}= \left\{ \begin{array}{lll}
     \textbf{u}_{1,j,k,l,t}=\bm{\nu}_{j,k,l,t}-\bm{\nu}_{j-1,k,l,t}  \\
     \textbf{u}_{2,j,k,l,t}=\bm{\nu}_{j,k,l,t}-\bm{\nu}_{j,k-1,l,t} \\
     \textbf{u}_{3,j,k,l,t}=\bm{\nu}_{j,k,l,t}-\bm{\nu}_{j,k,l-1,t}  \\
     \textbf{u}_{4,j,k,l,t}=\gamma(\bm{\nu}_{j,k,l,t}-\bm{\nu}_{j,k,l,t-1}) 
\end{array}\right.
\label{eq:diffindyn}
\end{equation*}
where $\gamma \in \mathbb{R}$ is the multiplicative factor along the time dimension. On the other hand, $\Tilde{F}_1$ and $\Tilde{F}_2$ are functions define as,
\begin{equation}
 \Tilde{F}_1( \tilde{\textbf{y}}_1)=\beta_1\Vert \tilde{\textbf{y}}_1\Vert_{2,1} \quad \text{and} \quad \Tilde{F}_2( \tilde{\textbf{y}}_2)=\beta_2\Vert \tilde{\textbf{y}}_2\Vert_{2,1}
\label{eq:mixednormdyn}
\end{equation}
where $\beta_1 \in \mathbb{R}$ and $\beta_2 \in \mathbb{R}$ are regularization hyper-parameters, and $\Vert .\Vert_{2,1}$ is the mixed (2,1)-norm defined in \eqref{eq:mixednormdef}. As in the static one-step method presented above, we design regularization for our specific dynamic application, i.e to be able to track the flow of iodinated contrast agent through the brain vasculature from a single dual energy CBCT scan. We include a sparsity constraint on the iodine image which contains the blood vessels with sparse structures and we use non-negativity constraints on water and iodine images as in the static one-step method. The new constraint which is specific to the dynamical method is a constraint which defines a set $\Omega$ of voxels from the 4D iodine images which are static along the time dimension. Typically the voxels of the skull or the background must be static.  Therefore for the dynamic one-step method we define two functions $\Tilde{G}_1$ and $\Tilde{G}_2$ expressed as $\Tilde{G}_1(\tilde{\textbf{x}})=\mathcal{X}_{\geq \mathbf{0}}\left(\textbf{x}_w\right)$ and $\Tilde{G}_2(\tilde{\textbf{x}})=\beta_3 \Vert \textbf{x}_{i,.}\Vert_1 + \mathcal{X}_{\geq \mathbf{0}}\left(\textbf{x}_{i,.}\right) + \textbf{I}_\Omega\left(\textbf{x}_{i,.}\right)$ where $\mathcal{X}_{\geq \mathbf{0}}$ is the indicator function defined above \eqref{eq:indicatorpositivity}, $\beta_3 \in \mathbb{R}$ is a regularization hyper-parameter, and $\textbf{I}_\Omega\left(.\right)$ is the term which introduces the static mask constraint on the 4D iodine image.
 Let $\Omega$ as the set of the voxels included in the static mask, the function $\textbf{I}_\Omega\left(.\right)$ can be defined as
\begin{equation*}
\textbf{I}_\Omega(\bm{\nu}) = \left\{ \begin{array}{ll}
0 \hspace{1mm} & \textbf{if}~\forall~\bm{\nu}_{j,k,l,t} \in  \Omega,   ~\bm{\nu}_{j,k,l,t}=\sum_{t=1}^T \bm{\nu}_{j,k,l,t}/T \vspace{0.5mm}, \\
+\infty &\text{otherwise.}
\end{array} \right.
\end{equation*}
Finally, the  proposed  dynamic  one-step method can be expressed as the following optimization problem,
\begin{equation}
  \hat{\tilde{\textbf{x}}} \in  \argmi{\tilde{\textbf{x}}}~ \sum_{h=0}^2 \Tilde{F}_h\left(\Tilde{K}_h(\tilde{\textbf{x}})\right) + \sum_{\xi=1}^2 \Tilde{G}_\xi(\tilde{\textbf{x}})
  \label{eq:dynamicoptimpb}
\end{equation}
Equivalently, we can reformulate the non-linear optimization problem \eqref{eq:dynamicoptimpb} as the following primal-dual formulation,
\begin{equation}
    \hat{\tilde{\textbf{x}}} \in  \argmi{\tilde{\textbf{x}}}\sum_{h=0}^2 \underset{\tilde{\textbf{y}}_h}{\text{max}} \langle  \Tilde{K}_h(\tilde{\textbf{x}}),\tilde{\textbf{y}}_h\rangle - \Tilde{\textit{F}}^*_h\left(\tilde{\textbf{y}}_h\right) + \sum_{\xi=1}^2 \Tilde{G}_\xi(\tilde{\textbf{x}})
  \label{eq:minmaxpbdyn}
\end{equation}
where $\tilde{\textbf{y}}_0 \in \mathbb{R}^{2M}$, $\Tilde{\textbf{y}}_1 \in \mathbb{R}^{3N}$, $\Tilde{\textbf{y}}_2 \in \mathbb{R}^{4TN}$ are the dual variables and for all h=\{0,1,2\} the function $\Tilde{\textit{F}}_h^*$ is the convex conjugate of the function $\Tilde{\textit{F}}_h$.\\
As in the static case, we use the Exact NL-PDHGM framework \cite{valkonen2014primal} to find a saddle point of the primal-dual optimization problem~\eqref{eq:minmaxpbdyn}, we explain how we have adapted the Exact NL-PDHGM algorithm to our dynamic one-step method. A parrallel is done with the associated pseudo-code Alg.\ref{algo1}.\\
$\bullet$ Eq. $y^{n+1}=\left(I +\sigma \partial F^*\right)^{-1}\left(y^n+ \sigma K(x^n_\omega) \right)$ (2.4c) :
\begin{equation*}
    \left|
    \begin{array}{ll}
       \tilde{\textbf{y}}^{n+1}_0=\left(I +\sigma_0 \partial \Tilde{F}_0^*\right)^{-1}\left(\tilde{\textbf{y}}^n_0+ \sigma_0 \Tilde{K}_0(\bar{\textbf{x}}^n) \right) \hspace{2mm} \text{(Alg.\ref{algo1}-l.\ref{grad0dyn1}-\ref{grad0dyn2})}  \\ \\
       \tilde{\textbf{y}}^{n+1}_1=\left(I +\sigma_1 \partial \Tilde{F}_1^*\right)^{-1}\left(\tilde{\textbf{y}}^n_1+ \sigma_1 \tilde{K}_1(\bar{\textbf{x}}^n) \right)  \quad \text{(Alg.\ref{algo1}-l.\ref{projball1dyn})}\\ \\
       \tilde{\textbf{y}}^{n+1}_2=\left(I +\sigma_2 \partial \Tilde{F}_2^*\right)^{-1}\left(\tilde{\textbf{y}}^n_2+ \sigma_2 \Tilde{K}_2(\bar{\textbf{x}}^n) \right)  \quad \text{(Alg.\ref{algo1}-l.\ref{projball2dyn})}
    \end{array}\right.
\end{equation*}
$\bullet$ Eq. $x^{n+1}=\left(I +\tau \partial G \right)^{-1}\left(x^n- \tau [\nabla K(x^n_\omega)]^*y^{n+1} \right)$ (2.4a)~:
\begin{equation*}
  \left|
    \begin{array}{ll}
    \tilde{\textbf{x}}^{n+1}_w=\left(I +\tau \partial \Tilde{G}_1 \right)^{-1}\left(\tilde{\textbf{x}}^n_w- \tau (\tilde{\textbf{v}}_{0,w}^{n+1}+\tilde{\textbf{v}}_1^{n+1})\right)~\text{(Alg.\ref{algo1}-\ref{proxG1dyn})}\\
    \tilde{\textbf{x}}^{n+1}_{i,.}=\left(I +\tau \partial \Tilde{G}_2 \right)^{-1}\left(\tilde{\textbf{x}}^n_{i,.}- \tau (\tilde{\textbf{v}}_{0,i,.}^{n+1}+\tilde{\textbf{v}}_2^{n+1})\right)~\text{(Alg.\ref{algo1}-l.\ref{proxG2dyn})}
    \end{array}\right.
\end{equation*}
where $\tilde{\textbf{v}}_1^{n+1}=\Tilde{K}_1^*\tilde{\textbf{y}}^{n+1}_1$, $\tilde{\textbf{v}}_2^{n+1}=\Tilde{K}_2^*\tilde{\textbf{y}}^{n+1}_2$ and $\tilde{\textbf{v}}_0^{n+1}=[\nabla\Tilde{K}_0(\bar{\textbf{x}}^n)]^*\tilde{\textbf{y}}_0^{n+1}$ with with $\Tilde{K}_1^*$, $\Tilde{K}_2^*$ and $[\nabla\Tilde{K}_0(\bar{\textbf{x}}^n)]^*$ the adjoint operators of $\Tilde{K}_1$, $\Tilde{K}_2$ and $[\nabla\Tilde{K}_0(\bar{\textbf{x}}^n)]$. The expressions of $\tilde{\textbf{v}}_1^{n+1}$ and $\tilde{\textbf{v}}_2^{n+1}$ are similar to those of the static case (see Alg.\ref{algo1}-l.\ref{div1dyn}-\ref{div2dyn}), except that the temporal component of 4D divergence operator $\operatorname{div}_{4D}^\gamma$ is multiplied by a factor $\gamma$. On the other hand, the expression of $\tilde{v}_0^{n+1}$ (Alg.\ref{algo1}-l.\ref{grad1dyn}-\ref{grad2dyn}) is non-trivial due to the non-linearity of the operator $\Tilde{K}_0$. As in the static case, the Appendix gives a proof of the expression of $\tilde{\textbf{v}}_0^{n+1}$.\\

$\bullet$ Eq. $x^{n+1}_\omega=x^{n+1}+\omega (x^{n+1}-x^n)$ (2.4b)  :
\begin{equation*}
\left|
    \begin{array}{ll}
     \bar{\textbf{x}}^{n+1}_w = \textbf{x}_w^{n+1} + \omega \left( \tilde{\textbf{x}}_w^{n+1} - \tilde{\textbf{x}}_w^{n}\right) \quad \text{(Alg.\ref{algo1}-l.\ref{omeg1dyn})}\\
     \bar{\textbf{x}}^{n+1}_{i,.} = \tilde{\textbf{x}}_{i,.}^{n+1} + \omega \left( \tilde{\textbf{x}}_{i,.}^{n+1} - \tilde{\textbf{x}}_{i,.}^n\right) \quad \text{(Alg.\ref{algo1}-l.\ref{omeg2dyn})}
    \end{array}\right.
\end{equation*}

\begin{algorithm}\label{algo1}
 \caption{The dynamical one-step algorithm}
\SetAlgoLined
 Initialize all variables, choose $\omega \in [0, 1]$ and $\tau_h$, $\sigma\geq 0$ such that $\tau \sigma_h \Vert \Tilde{K}_h\Vert^2 <1$ \;
 \For{n = 0 to niter-1}{
     $\tilde{\textbf{y}}_{0,1}^{n+1} = \dfrac{2}{2+\sigma_0}\left( \tilde{\textbf{y}}_{0,1}^{n}+\sigma_0 (\Tilde{\textbf{m}}_1(\textbf{A}\bar{\textbf{x}}_w^n,\Tilde{\textbf{A}}\bar{\textbf{x}}_{i,.}^n)- \textbf{s}_{1} )\right)$ \label{grad0dyn1}\\
     $\tilde{\textbf{y}}_{0,2}^{n+1} = \dfrac{2}{2+\sigma_0}\left( \tilde{\textbf{y}}_{0,2}^{n}+\sigma_0 (\Tilde{\textbf{m}}_2(\textbf{A}\bar{\textbf{x}}_w^n,\Tilde{\textbf{A}}\bar{\textbf{x}}_{i,.}^n)- \textbf{s}_{2} )\right)$  \label{grad0dyn2}\\ 
     $\tilde{\textbf{y}}_{1}^{n+1}=proj_{\beta_1 P} \left(\tilde{\textbf{y}}_{1}^{n} +\sigma_1 \nabla_{3D}\bar{\textbf{x}}_w^n \right)$ \label{projball1dyn}\\
     $\tilde{\textbf{y}}_{2}^{n+1}=proj_{\beta_2 P} \left(\tilde{\textbf{y}}_{2}^{n} +\sigma_2 \nabla_{4D}^\gamma\bar{\textbf{x}}_{i,.}^n \label{projball2dyn}\right)$ \\    
      $\tilde{\textbf{v}}_{0,w}^{n+1} =\textbf{A}^\mathsf{T}(\underset{c=1}{\overset{2}{\sum}}(2a_{5c}\textbf{A}\bar{\textbf{x}}_w^n+a_{3c}\Tilde{\textbf{A}}\bar{\textbf{x}}_{i,.}^n+a_{2c}\textbf{1})\odot \tilde{\textbf{y}}_{0,c}^{n+1})$ \label{grad1dyn}\\
      $\tilde{\textbf{v}}_{0,i,.}^{n+1} = \Tilde{\textbf{A}}^\mathsf{T}(\underset{c=1}{\overset{2}{\sum}}(2a_{4c}\Tilde{\textbf{A}}\bar{\textbf{x}}_{i,.}^n+a_{3c}\textbf{A}\bar{\textbf{x}}_w^n+a_{1c}\textbf{1})\odot \tilde{\textbf{y}}_{0,c}^{n+1})$\label{grad2dyn}\\    
     $\tilde{\textbf{v}_{1}}^{n+1}=-\operatorname{div}\left( \tilde{\textbf{y}}_{1}^{n+1}\right)$\label{div1dyn}\\
     $\tilde{\textbf{v}_{2}}^{n+1}=-\operatorname{div}_{4D}^\gamma\left(\tilde{\textbf{y}}_{2}^{n+1}\right)$\label{div2dyn}\\
     $\tilde{\textbf{x}}_w^{n+1}=proj_{\mathbb{R}^{N+}}\left(\tilde{\textbf{x}}_w^{n}-\tau \tilde{\textbf{v}}_{0,w}^{n+1}-\tau \tilde{\textbf{v}}_{1}^{n+1}\right)$ \label{proxG1dyn}\\
     $\tilde{\textbf{x}}_{i,.}^{n+1}=prox_{\textbf{I}_\Omega}\left(\textit{S}^+_{\tau\beta_3}\left(\tilde{\textbf{x}}_{i,.}^n-\tau \tilde{\textbf{v}}_{0,i,.}^{n+1}-\tau \tilde{\textbf{v}}_{2}^{n+1}\right)\right)$ \label{proxG2dyn}\\
     $\bar{\textbf{x}}^{n+1}_w = \tilde{\textbf{x}}_w^{n+1} + \omega \left( \tilde{\textbf{x}}_w^{n+1} - \tilde{\textbf{x}}_w^{n}\right)$ \label{omeg1dyn}\\
     $\textbf{for}~\textit{t=1 to T}~\textbf{do}~\bar{\textbf{x}}^{n+1}_{i,t} = \tilde{\textbf{x}}_{i,t}^{n+1} + \omega \left( \tilde{\textbf{x}}_{i,t}^{n+1} - \tilde{\textbf{x}}_{i,t}^{n}\right)$~\textbf{end} \label{omeg2dyn}
}
\end{algorithm}

In line \ref{proxG2dyn} the proximity operator $prox_{\textbf{I}_\Omega}$ of the indicator function $\textbf{I}_\Omega$ is defined by :
\begin{equation*}
prox_{\textbf{I}_\Omega}\left(\bm{\nu}\right)_{j,k,l,t} = \left\{ \begin{array}{ll}
&\sum_{t=1}^T \bm{\nu}_{j,k,l,t}/T \hspace{2mm} \textbf{if} \quad \bm{\nu}_{j,k,l,t}\in \Omega  \vspace{0.5mm} \\
&\bm{\nu}_{j,k,l,t} \hspace{15mm} \textbf{if} \quad \bm{\nu}_{j,k,l,t} \notin \Omega.
\end{array} \right.
\label{eq:proxsoft}
\end{equation*}

\section{Experiments}
\subsection{Simulations}

Data were simulated to produce acquisitions from a dual-energy CBCT system with a flat-panel detector, with a medical application focused on the brain imaging in stroke. For these simulations we consider a C-arm architecture with 620 projections acquired over 205 degrees in 25 seconds, with a source voltage of 120\,kV and the tube load was set to 1.25\,mAs. The simulated CBCT system uses a 2D detector of 198$\times$256 pixels with a 1.48\,mm pitch, the distance between source and detector is 1195\,mm and the distance between object and detector is 390\,mm.
\begin{figure}[htb]
\centering
\includegraphics[width=3.8cm]{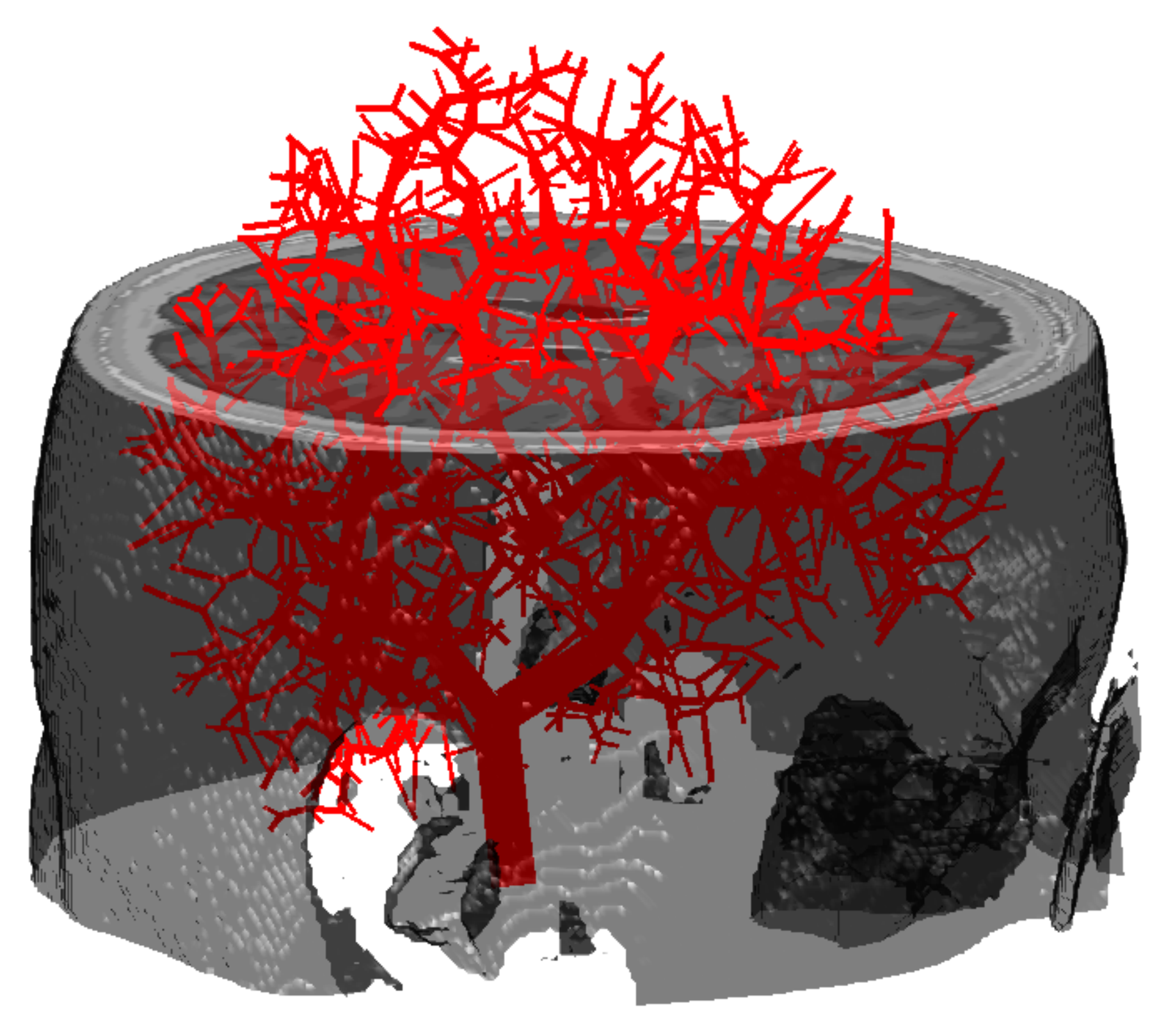}
\caption{ \label{fig:Phantom} Brain phantom with vascular tree.}
\end{figure}

In \cite{sones1989noise} the authors showed that the noise between both layers is uncorrelated. That is why a photonic poissonian noise is added for each dual-energy measurement before the log transform was applied. The forward projection was obtained with a dual energy CBCT simulator which used a forward projector different from the one used for the reconstruction. The energy spectrum for each detector layer was simulated with a discretization of 1 keV from 12 keV to 150 keV.\\ While we consider the same system and acquisition parameters for all simulations, we used two different versions of a brain phantom (see Fig.\ref{fig:Phantom}) for both simulations : the first one considers a head phantom with a static vascular tree, i.e. without evolution of the iodine concentration in the blood vessels during the scan, whereas the second simulation considers a head phantom with a dynamic vascular tree. The detail of the brainweb phantom used (without vascular tree) for these simulations is available in \cite{simon2020physical}.

\subsubsection{Simulation I : A static brain phantom} \label{sec:simustatic}
The static part of this brain phantom is based on the brainweb phantom \cite{cocosco1997brainweb}, a voxelized head phantom composed by 10 different tissue classes. To this software phantom, a vascular tree has been added as described in \cite{heylen20194d}. The vascular tree has an iodine concentration of 20 mg/mL. A rendering of the tree is shown in figure \ref{fig:Phantom}. 
\subsubsection{Simulation II : A dynamic brain phantom}  \label{sec:simudynamic}
For the dynamic part, the time dependent iodine concentration within the vascular tree was computed with a dynamic model, as proposed in \cite{heylen20194d}. The vascular tree has a wide initial artery segment low in the brain, and generation of arterial output terminals was constrained to gray/white matter tissue classes. Each tube segment was considered to possess laminar flow so that dispersion and time delays can be calculated analytically \cite{taylor1953dispersion}, therefore a realistic time behaviour of the flow of contrast is obtained. No draining venous network was simulated, and the contrast will disappear at the arterial output terminals of the vascular tree. The artificial vascular tree was voxelized, and added to the brainweb phantom as an additional dynamic class. For each projection in the sinogram, an appropriate time point was calculated, and the corresponding iodine contribution from the vascular tree determined.

\section{Results} \label{sec:Resultstatic}
\subsection{Results on the static phantom}
\begin{figure*}
\centering
\includegraphics[width=16cm]{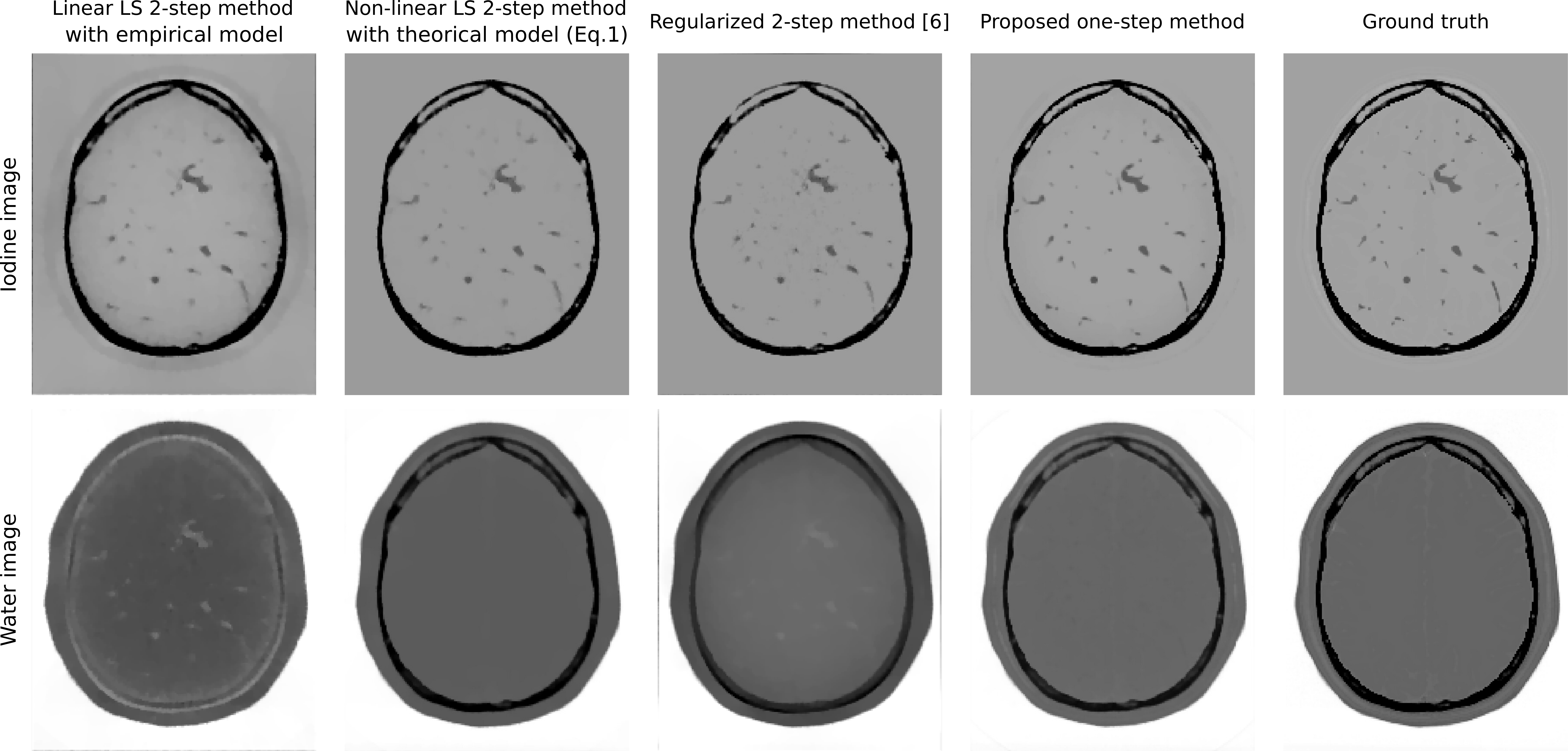}
\caption{ \label{fig:Results_static} Comparison of iodine images (first row) and water images (second row) obtained with the \textbf{Method 1} (first column), the \textbf{Method 2} (second column), the \textbf{Method 3} (third column)  and the proposed static one-step method (fourth column). The fifth column is the iodine ground truth.}
\label{fig:static}
\end{figure*}
To evaluate the static one-step method which is proposed in Sec.\ref{Sec:staticonestep} we used the simulation I and three other 2-step methods. \textbf{Method 1} is a 2-step method which first computes multi-material projections with a non-iterative maximum likelihood method using an empirical model and then applies a Total Variation constrained tomographic reconstruction. 
\textbf{Method 2} is a 2-step method which first computes multi-material projections with a non-linear iterative maximum likelihood method using the theorical model \eqref{eq:modelth} and then applies a Total Variation constrained tomographic reconstruction. This method is the method described in \cite{heylen20194d} with a single time frame. 
\textbf{Method 3} is a 2-step method which first computes multi-material projections with a regularized decomposition method including sparsity constraints, a Tikhonov regularization and a non-negativity constraint \cite{jolivet2021twostep} and then a Total Variation constrained tomographic reconstruction. Method 1 and method 3 use the empirical model used in \cite{jolivetIEEE2020}. In figures \ref{fig:Results_static} method 1 is called "LS 2-step method with empirical model", method 2 is called "LS 2-step method with theorical model" whereas the method 3 is called "Regularized 2-step method".
Figure \ref{fig:Results_static} shows a comparison of the reconstructions obtained with the four methods from the simulation I. All methods reconstructed a voxelized object of  $181\times 217\times 181$ voxels with a 1 $\text{mm}^3$ voxel size. Second step of the methods 1, 2 and 3 use a Total variation constrained tomographic reconstruction \cite{sidky2012convex} using 500 iterations, whereas the proposed one-step method used 500 iterations.

The comparison shows that the unconstrained method (method 1) is sensible to the data noise and gives anti-correlated artifacts which introduce a material crosstalk between iodine and water images. The noise amplification due to the ill-conditioning of the inversion step in the basis change and the material crosstalk effects were significantly reduced by the non-negativity constraint and regularization introduced in methods 2 and 3. Compared to the two-step methods, the proposed one-step method gives the best results in terms of decomposition and signal-to-noise ratio of reconstructions.
Table \ref{RMSEstatic} shows quantitative comparisons where we used the Root-Mean-Square Error (RMSE) criterion to evaluate the difference between material reconstructions from different methods and the ground truths. The first row is a comparison from the water map, the second row is a comparison from the iodine map excluding voxels associated to the skull whereas the third row is a comparison from the iodine map and ground truth only on voxels associated to the blood vessels. Table \ref{RMSEstatic} shows results for each method which confirm that the proposed one-step method produced the best results.

\begin{table}[!ht]
\centering
\begin{tabular}{c|c|c|c|c}
  RMSE & Method 1 & Method 2 & Method 3 & Proposed\\
  \hline
  water  & $0.15$ & $0.11$ & $0.12$  & $0.10$ \\
  \hline
  iodine & $4\times 10^{-4}$ & $9\times 10^{-5}$ & $1.0\times 10^{-4}$ & $5\times 10^{-5}$ \\
  \hline
  iodine & $2.3\times 10^{-4}$ & $1.4\times 10^{-4}$ & $1.4\times 10^{-4}$  & $7.5\times 10^{-5}$ \\
  \tiny{(blood vessels)} &  & & &

\end{tabular} \caption{rmse between reconstructions and ground truth.}
\label{RMSEstatic}
\end{table}

\subsection{Results on the dynamic phantom} \label{sec:Resultdynamic}
\begin{figure*}
\centering
\includegraphics[width=16cm]{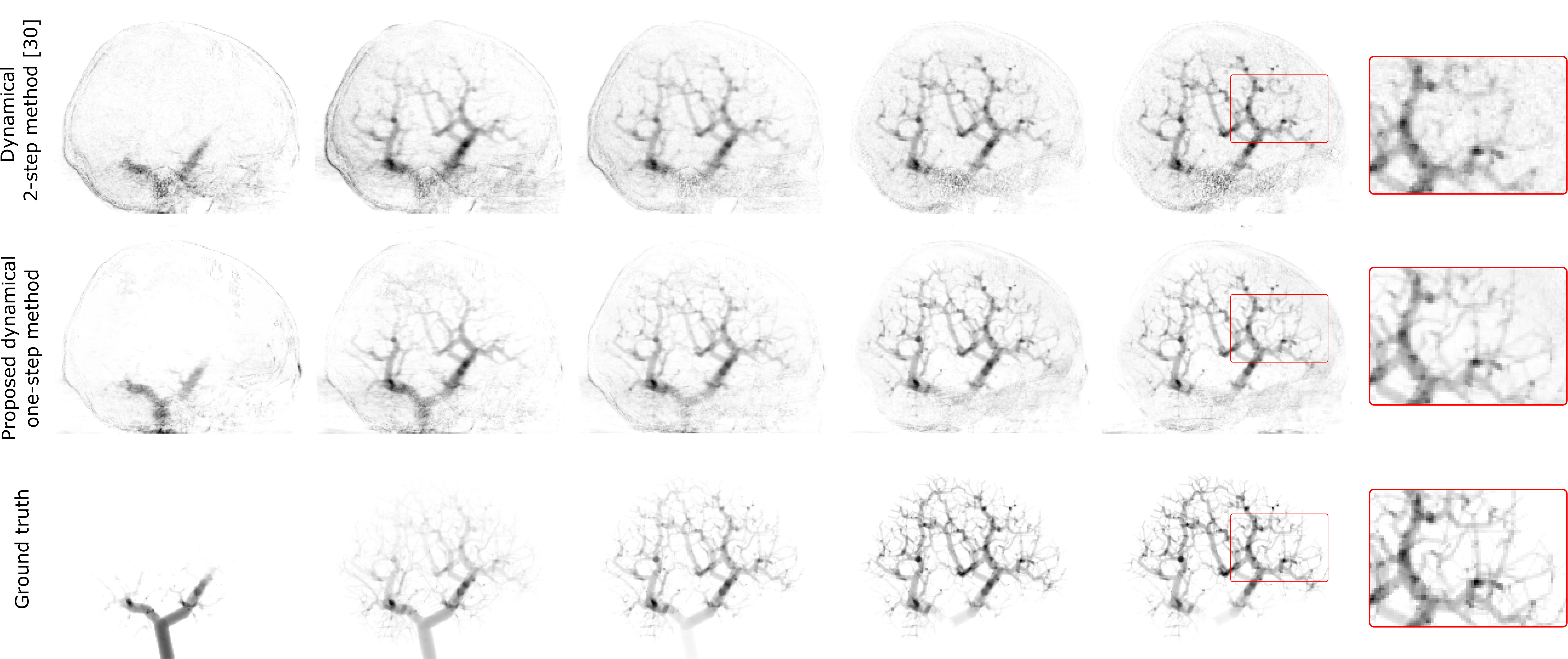}
\caption{ \label{fig:Resultats} MIP visualization of dynamical iodine reconstructions obtained with the dynamical 2-step method \cite{heylen20194d} (first row) and the proposed dynamical one-step method (second row), and the ground truth (third row).}
\end{figure*}

To evaluate the dynamic one-step method which is proposed in Sec.\ref{Sec:dynamiconestep} we used the simulation II and another published dynamical method similar to \cite{heylen20194d}. The method \cite{heylen20194d} is a dynamical 2-step method, which applies a material decomposition followed by a constrained dynamical tomographic reconstruction with a 4D Total-Variation regularization from the iodine projections. In this study, the interest of the dynamical reconstruction is to track the evolution of the iodine concentration in the blood vessels. That is why in both methods, we reconstruct 10 times frames (T=10) of 181$\times$217$\times$181 voxels with a 1 $\text{mm}^3$ voxel size. In both methods, we use the same static mask, obtained from a thresholding on a combination between water and iodine static reconstructions, to define the set $\Omega$. Both methods are computed with 200 iterations. Fig.\ref{fig:Resultats} shows only 5 digital subtractions between times frames and the first time frame (one in two) of the iodine concentration reconstructed for the dynamical 2-step method (first row) and the proposed dynamical one-step method (second row). The third row is the iodine concentration ground truth at different time points. As in the static case, we can see that the dynamical one-step method obtained a better signal-to-noise ratio.  
Table \ref{RMSEdynamic} shows quantitative comparisons where we used a the Root-Mean-Square Error (RMSE) criterion to evaluate the difference between iodine reconstructions substracted to the first time reconstruction and the ground truths. The first row is a comparison excluding voxels associated to the skull whereas the second row is a comparison using only voxels associated to the blood vessels. Table \ref{RMSEdynamic} shows that the proposed one-step method produced better results than the dynamic 2-step method \cite{heylen20194d}.

\begin{table}[!ht]
\centering
\begin{tabular}{c|c|c}
  RMSE & Dynamic 2-step & Proposed dynamic one-step\\
  \hline
  iodine  & $8.39\times 10^{-5}$ & $5.85\times 10^{-5}$ \\
  \hline
  iodine & $5.3\times 10^{-4}$ & $3.1\times 10^{-4}$ \\
    \tiny{(blood vessels)} &  &  \\
\end{tabular} \caption{rmse between reconstructions and ground truth.}
\label{RMSEdynamic}
\end{table}

\section{Discussions} \label{sec:discussion}
\subsection{Initialization, computational time and convergence...}
\begin{figure}[h]
\centering
\includegraphics[width=8.7cm]{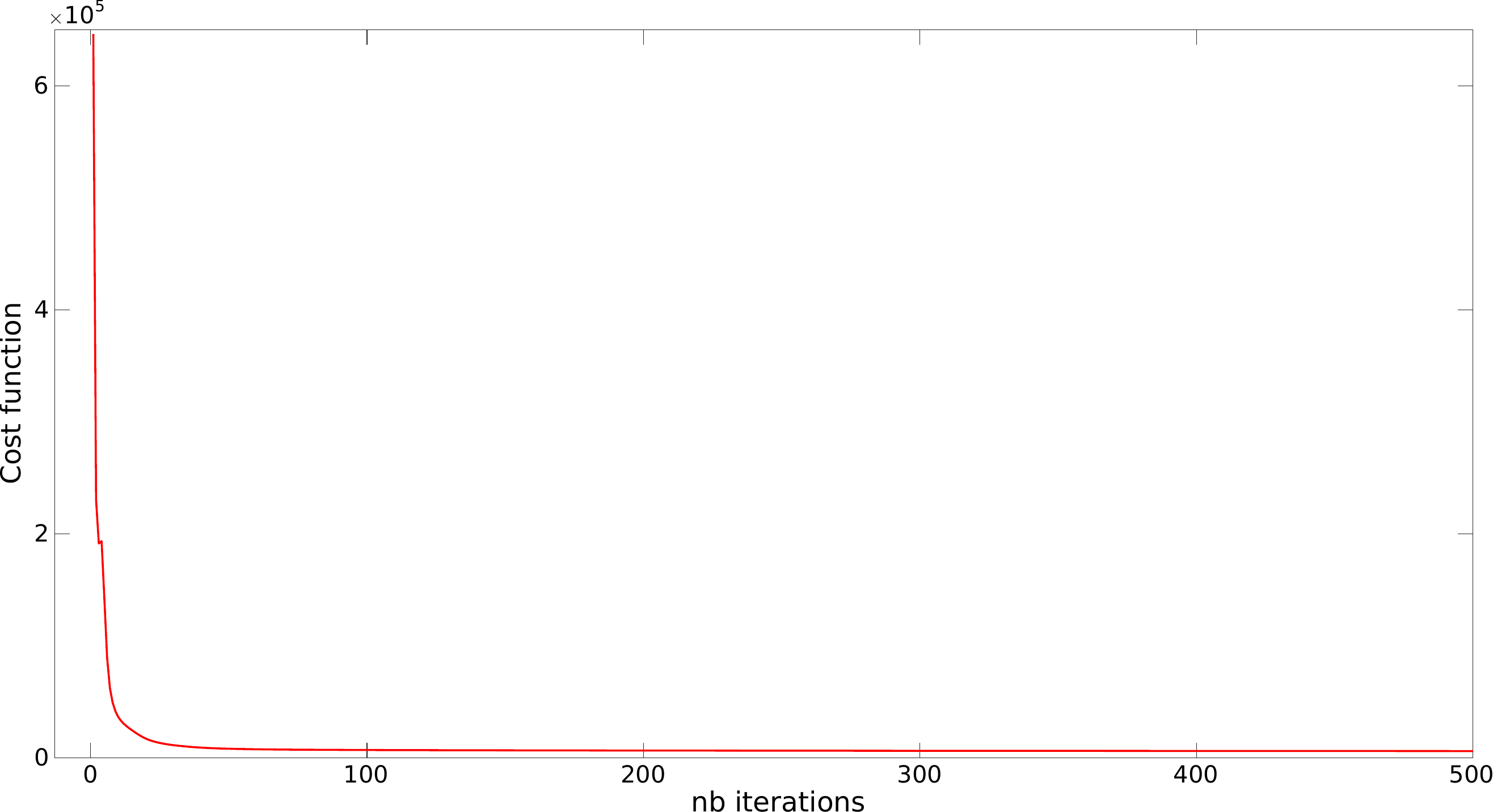}
\caption{ \label{fig:cost_fun} Evolution of the cost function.}
\end{figure}

While the static 2-step methods are initialized with null images, a good initialization is crucial to drastically reduce the number of iterations for convergence for the one-step methods. In this study static and dynamic methods are initialized with reconstructions obtained from the static 2-step method "Method 2". Using this initialization strategy for the one-step methods, some hundreds of iterations are enough to reach convergence (see Fig.\ref{fig:cost_fun}). Practically, we can note that $\Vert \textbf{s} \Vert_2^2$ (equivalent to the cost of data fidelity term with null material images) is around $4 \times 10^8$, whereas after 200 iterations the cost function is around $6 \times 10^3$. A GPU implementation is another key to reduce the time of calculation. In our case with a GPU NVIDIA TITAN XP, 200 iterations of the dynamic one-step method (with 10 time frames) run in approximately 1 hour whereas they run in approximately 30 minutes for the dynamical 2-step method. From an optimization point of view, the one-step methods solve a possibly non-convex and non-smooth optimization problem and converge to a critical point which is a local minimum but without guarantee to be the global minimum \cite{valkonen2014primal}.

\subsection{Tuning of hyper-parameters and parameters of the optimization algorithm}
For optimization-based CT image reconstruction the tuning of hyper-parameters is a relevant question \cite{bengio2000gradient,shen2018intelligent,xu2021patient}. In our application we have to set the parameters associated to the Valkonen optimization algorithm ($\tau$ and $\sigma_h$) and hyper-parameters giving a weight to each regularization function. In our case we have observed that the parameters $\tau$ and $\sigma_h$ are robust if we respect the condition $\tau \sigma_h \Vert K_h\Vert^2 <1$. In practice we normalized operators $K_h$ such as $\Vert K_h\Vert^2 = 1$. For the non-linear operator $K_0$ we pratically approximate its normalization with a normalization of the tomographic projector such as $\Vert\textbf{A}\Vert^2 =1$ and $\Vert\Tilde{\textbf{A}}\Vert^2 =1$. Therefore, we applied the condition $\tau \sigma_h < 1$ and more precisely we use $\tau=3.5$ and $\sigma_0=\sigma_1=\sigma_2=0.95/\tau$. 
On the other hand, the hyper-parameters can have more influence on the final results and are tuned manually. A good criterion to set automatically the hyper-parameters stays an open question, even if we can think reasonably that data acquired in the same conditions (same CBCT system, same medical protocol...) could lead to the similar hyper-parameters. 

\subsection{Static mask and number of time frames}
A critical point for a good dynamical reconstruction is to have a good static mask because it prevents dynamic behaviour in voxels known to be static and it reduces the number of unknows significantly, improving the conditioning of this reconstruction problem. A perspective will be to optimize this mask to have an accurate estimation of static voxels (typically the skull is static).

\section{Conclusion \& Perspectives} \label{sec:conclusion}
We have proposed static and dynamical one-step methods for dual-energy CT including sparsity constraints and based on the optimization strategy NL-PDHGM \cite{valkonen2014primal}. Using a dual energy CBCT simulation obtained from brain phantoms, we compare the proposed one-step methods with methods which were previously published. Simulation data used parameters of a C-arm DECT system close to the prototype described in \cite{staahl2021performance}. Therefore these promising results could probably be validated on clinical data obtained with this prototype. In future work, other regularization terms could be considered, such as the Directional-TV regularization which seems to give good results for limited angle data \cite{zhang2021directional,chen2021dual}. For clinical application, it will be important to optimize the calculation time to reduce the time of the medical diagnosis. Note that an extension for data from energy-resolved photon counting detectors \cite{taguchi2013vision} which could have more than 2 energy bins is straightforward. In this case a small modification of the proposed algorithms can lead to a decomposition of more than 2 materials.   

\section*{Acknowledgements}
This work was done under the NEXIS project, that has received funding from the European Union’s Horizon 2020 Research and Innovations Program (Grant Agreement no.780026). The authors are grateful to Klaus Jürgen Engel, Bernd Menser and Matthias Simon from Philips Research for providing the simulated used in this work and for many helpful discussions. The authors are grateful to Fredrik St\aa hl from the Karolinska Institutet/Karolinska University Hospital for many helpful discussions.

\appendices
\section{Expression of the proximity operator $\textbf{prox}_{\sigma_0 F_0^*}$}

In this part of the appendix, we give an expression of the conjugate of the function $F_0(.)$ (and equivalently of the function $\Tilde{F}_0(.)$)~:
\begin{equation}
    F_0^*(y)=\underset{\textbf{z}}{sup}~\langle \textbf{y},\textbf{z}\rangle - F_0(\textbf{z})
    = \underset{\textbf{z}}{sup}~\langle \textbf{y},\textbf{z}\rangle - \Vert \textbf{z} - \textbf{s}\Vert_2^2
    \label{F0star}
\end{equation}
Taking derivatives along $\textbf{z}$ and setting them to zero produces the supremum value $\hat{\textbf{z}}=\frac{\textbf{y}}{2}+\textbf{s}$, then substitution in \eqref{F0star} gives,
\begin{equation}
    F_0^*(\textbf{y})=\langle \textbf{y},\frac{1}{4}\textbf{y}+\textbf{s}\rangle
\end{equation}
Therefore, the proximity operator $\textbf{prox}_{\sigma_0 F_0^*}$ can be expressed as
\begin{eqnarray*}
    \left(I +\sigma_0 \partial F_0^*\right)^{-1}\left(\bm{\nu} \right)&=&\argmi{\textbf{z}} \dfrac{\Vert \textbf{z}-\bm{\nu}\Vert_2^2}{2\sigma_0} + F_0^*(\textbf{z})\\
    &=&\argmi{\textbf{z}} \dfrac{\Vert \textbf{z}-\bm{\nu}\Vert_2^2}{2\sigma_0} + \langle \textbf{z},\frac{1}{4}\textbf{z}+\textbf{s}\rangle\\
    &=&\frac{2}{2+\sigma_0}\left(\bm{\nu} - \sigma_0 \textbf{s}\right)~.\qquad \qquad \quad_\blacksquare
\end{eqnarray*}

\section{Expression of the proximity operator of the conjugate function of the (2,1)-mixed norm}
In this part of the appendix, we give an expression of the conjugate of the function $F_1(\textbf{x})=\alpha_1\Vert \textbf{x}\Vert_{2,1}$ which is proportional to the (2,1)-mixed norm. The expressions for $F_2(.)$, $\Tilde{F}_1(.)$ and $\Tilde{F}_2(.)$ are similar.\\

Let a function $F_1(\textbf{x})=\alpha_1\Vert \textbf{x}\Vert_{1,2}$, therefore its conjuguate function can be expressed as,
\begin{eqnarray}
    F_1^*(\textbf{x})&=&\underset{y\in \mathbb{R}^{JD}}{sup}~\langle \textbf{x},\textbf{y}\rangle -\alpha_1 F_1(\textbf{y}) \nonumber\\
    &=& \underset{\textbf{y}\in \mathbb{R}^{JD}}{sup}\sum_j \sum_d \textbf{x}_{j,d}\textbf{y}_{j,d}  -\alpha_1\sqrt{\sum_d \textbf{y}_{j,d}^2}\nonumber\\
    &=& \sum_j \underset{\textbf{y}_j\in \mathbb{R}^{D}}{sup} \sum_d \textbf{x}_{j,d}\textbf{y}_{j,d}  -\alpha_1\sqrt{\sum_d \textbf{y}_{j,d}^2} \label{eq:F1star}
\end{eqnarray}
\noindent where the supremum values $\hat{\textbf{y}}_j$ can be found setting the derivative to 0,
\begin{equation}
\textbf{x}_{j,d}  -\alpha_1\dfrac{\hat{\textbf{y}}_{j,d}}{\sqrt{\sum_d \hat{\textbf{y}}_{j,d}^2}}=0~.
    \label{eq:egalite_grad}
\end{equation}
Defining $\zeta_j=\Vert \hat{\textbf{y}}_{j}\Vert_2$ and using \eqref{eq:egalite_grad} then,
\begin{equation}
\hat{\textbf{y}}_{j,d}  =\dfrac{\zeta_j \textbf{x}_{j,d}}{\alpha_1}~.
\label{eq:egalite_grad2}
\end{equation}
Therefore we can reformulate \eqref{eq:F1star} as,
\begin{eqnarray}
F_1^*(\textbf{x})&=& \sum_j \underset{\zeta_j\in \mathbb{R}^+}{sup} \zeta_j\left(\sum_d \dfrac{\textbf{x}_{j,d}^2}{\alpha_1} -\alpha_1\right)\\
&=& \sum_j \underset{\zeta_j\in \mathbb{R}^+}{sup} \zeta_j\left(\dfrac{\Vert \textbf{x}_{j}\Vert^2_2}{\alpha_1} -\alpha_1\right)
\label{eq:egalite_grad3}
\end{eqnarray}
In \eqref{eq:egalite_grad3} supremum values can be expressed as
\begin{equation}
    \hat{\zeta}_j=\left\{ \begin{array}{ll}
         0 \quad &\text{if}\quad \Vert \textbf{x}_j\Vert_2 \leq \alpha_1\\
         +\infty  \quad &\text{if}\quad\Vert \textbf{x}_j\Vert_2 > \alpha_1 .
    \end{array}\right.
    \label{eq:supremumvalue}
\end{equation}
Plugging supremum values $\hat{\zeta}_j$ in \eqref{eq:egalite_grad3} leads to
\begin{eqnarray}
       F_1^*(\textbf{x})&=&\sum_j \left(\sum_d \textbf{x}_{j,d}\dfrac{\hat{\zeta}_j \textbf{x}_{j,d}}{\alpha_1}\right) -\alpha_1 \hat{\zeta}_j\\
    &=& \sum_j \hat{\zeta}_j \left(\dfrac{\Vert \textbf{x}_j\Vert_2^2}{\alpha_1} -\alpha_1 \right)\\
    &=& \left\{ \begin{array}{ll}
         0 &\text{if}\quad \forall j \quad \Vert \textbf{x}_j\Vert_2 \leq \alpha_1\\
         +\infty  \quad &\text{otherwise.}
    \end{array} \right. \label{eq:conjF1}
\end{eqnarray}
Using \eqref{eq:conjF1} we can expressed the proximity operator~$\textbf{prox}_{\sigma_1 F_1^*}$~:

\begin{eqnarray}
    \left(I +\sigma_1 \partial F_1^*\right)^{-1}\left(\bm{\nu} \right) = \argmi{\textbf{z}} \dfrac{\Vert \textbf{z}-\bm{\nu}\Vert_2^2}{2\sigma_1} +  F_1^*(\textbf{z})
\end{eqnarray}
Therefore, 
\begin{eqnarray}
    \textbf{prox}_{\sigma_1 F_1^*}(\bm{\nu})_{j,d}= \left\{ \begin{array}{cl}
         \bm{\nu}_{j,d} &\text{if}\quad \Vert \bm{\nu}_j\Vert_2 \leq \alpha_1\vspace{1mm} \\
         \alpha_1 \dfrac{\bm{\nu}_{j,d}}{\Vert \bm{\nu}_j\Vert_2}  \quad &\text{otherwise.}
    \end{array} \right.
    \label{eq:finalproxmixed}
\end{eqnarray}
Using \eqref{eq:finalproxmixed} we can see that the proximity operator $\textbf{prox}_{\sigma_1 F_1^*}(\bm{\nu})$ can be solved with a projection of each element $\nu$ onto the $\ell_2$-ball of radius $\alpha_1$.$\hspace{35mm} \quad_\blacksquare$

\section{Proof of the expression $[\nabla K_0(\mathbf{x})]^*\mathbf{y}_0$} 

\label{AppendixA}
In this part of the appendix we give a proof of the expression of $[\nabla \Tilde{K}_0(\bar{\textbf{x}})]^*\tilde{\textbf{y}}_{0}^{n+1}$ (and transparently for $[\nabla K_0(\bar{\textbf{x}})]^*\textbf{y}_{0}^{n+1}$). To translate this explanation about $[\nabla \Tilde{K}_0(\bar{\textbf{x}})]^*\tilde{\textbf{y}}_{0}^{n+1}$ to $[\nabla K_0(\bar{\textbf{x}})]^*\textbf{y}_{0}^{n+1}$), change the operator  $\Tilde{K}_0$ to $K_0$, $\tilde{A}$ to $A$, and $\textbf{x}_{i,.}$ to $\textbf{x}_{i}$.\\

Let $\Tilde{K}(\bar{\textbf{x}})$ be the non-linear forward model defined as,
\begin{equation}
\Tilde{K}_0(\bar{\textbf{x}})=g\circ f (\bar{\textbf{x}})
\label{eq:eq1appendix}
\end{equation}
In our case,
\begin{equation*}
f (\bar{\textbf{x}})=\begin{pmatrix} \textbf{A} & \textbf{0} \\
\textbf{0} & \Tilde{\textbf{A}}
\end{pmatrix}
\underbrace{\begin{pmatrix} \bar{\textbf{x}}_w \\
\bar{\textbf{x}}_{i,.}
\end{pmatrix}}_{\bar{\textbf{x}}}\hspace{0.1mm} \text{and} \hspace{1mm} g(\textbf{l}_w,\textbf{l}_{i})=\begin{pmatrix} \Tilde{\textbf{m}}_{1}\left(\textbf{l}_w,\textbf{l}_{i}\right) \\
\Tilde{\textbf{m}}_{2}\left(\textbf{l}_w,\textbf{l}_{i}\right)
\end{pmatrix}.
\label{eq:eq2appendix}
\end{equation*}
Let $\textbf{l}=\begin{pmatrix}\textbf{l}_w\\ \textbf{l}_{i} \end{pmatrix}$, then $f(\bar{\textbf{x}})=\textbf{l}$.\\ \\
\vspace{5mm}
\underline{Jacobian matrix of $f$}
\begin{equation}
J_f(\bar{\textbf{x}})=\begin{pmatrix} \textbf{A} & \textbf{0} \\
\textbf{0} & \Tilde{\textbf{A}}
\end{pmatrix}
\label{eq:eq3appendix}
\end{equation}
\underline{Jacobian matrix of $g$}\\
Let $M$ the number of elements of each material sinogram. We define matrices \textbf{E}, \textbf{F}, \textbf{G}, \textbf{H} such that $\forall j \in \llbracket 1, M \rrbracket $ :
$$ \textbf{E}_{jj} = \dfrac{\partial \Tilde{\textbf{m}}_{1}\left(\textbf{l}_w,\textbf{l}_{i}\right)}{\partial \left(\textbf{l}_w\right)_j} = \left( 2a_{51}\textbf{l}_w + a_{31}\textbf{l}_{i}+a_{21}\textbf{1}\right)_j $$\\
$$ \textbf{F}_{jj} = \dfrac{\partial \Tilde{\textbf{m}}_{2}\left(\textbf{l}_w,\textbf{l}_{i}\right)}{\partial \left(\textbf{l}_w\right)_j} = \left( 2a_{52}\textbf{l}_w + a_{32}\textbf{l}_{i}+a_{22}\textbf{1}\right)_j $$\\
$$ \textbf{G}_{jj} = \dfrac{\partial \Tilde{\textbf{m}}_{1}\left(\textbf{l}_w,\textbf{l}_{i}\right)}{\partial \left(\textbf{l}_{i}\right)_j} = \left( 2a_{41}\textbf{l}_{i} + a_{31}\textbf{l}_w+a_{11}\textbf{1}\right)_j $$\\
$$ \textbf{H}_{jj} = \dfrac{\partial \Tilde{\textbf{m}}_{2}\left(\textbf{l}_w,\textbf{l}_{i}\right)}{\partial \left(\textbf{l}_{i}\right)_j} = \left( 2a_{42}\textbf{l}_{i} + a_{32}\textbf{l}_w+a_{12}\textbf{1}\right)_j $$
and $\forall~ k\neq j, ~\textbf{E}_{kj}= \textbf{F}_{kj}= \textbf{G}_{kj}= \textbf{H}_{kj}=0$. Then, 
\begin{equation}
J_g(\textbf{l})=\begin{pmatrix} \dfrac{\partial \Tilde{\textbf{m}}_{1}\left(\textbf{l}_w,\textbf{l}_{i}\right)}{\partial \textbf{l}_w} & \dfrac{\partial \Tilde{\textbf{m}}_{1}\left(\textbf{l}_w,\textbf{l}_{i}\right)}{\partial \textbf{l}_{i}} \\
\dfrac{\partial \Tilde{\textbf{m}}_{2}\left(\textbf{l}_w,\textbf{l}_{i}\right)}{\partial \textbf{l}_w} & \dfrac{\partial \Tilde{\textbf{m}}_{2}\left(\textbf{l}_w,\textbf{l}_{i}\right)}{\partial \textbf{l}_{i}}
\end{pmatrix}=\begin{pmatrix} \textbf{E} & \textbf{G}\\ \textbf{F} & \textbf{H}
\end{pmatrix}.
\label{eq:eq4appendix}
\end{equation}

\underline{Jacobian matrix of $g \circ f$}\\
The Jacobian matrix of $g \circ f$ can be written as,
\begin{equation}
    J_{g \circ f}(\bar{\textbf{x}}) = J_g(\textbf{l}) J_f(\bar{\textbf{x}}) 
\end{equation}
with results \eqref{eq:eq3appendix} and \eqref{eq:eq4appendix}, then
\begin{equation}
    J_{g \circ f}(\bar{\textbf{x}}) = \begin{pmatrix} \textbf{E} & \textbf{G}\\ \textbf{F} & \textbf{H}
\end{pmatrix} \begin{pmatrix} \textbf{A} & \textbf{0} \\
\textbf{0} & \Tilde{\textbf{A}}
\end{pmatrix}.
\end{equation}
In the Exact NL-PDHGM framework \cite{valkonen2014primal}, $\nabla \tilde{K}_0(\bar{\textbf{x}}) = J_{g \circ f}(\bar{\textbf{x}})$, then
\begin{eqnarray*}
 \left[\nabla\Tilde{K}_0(\bar{\textbf{x}}) \right]^*=\left[\nabla\Tilde{K}_0(\bar{\textbf{x}}) \right]^\mathsf{T}&=&\left[ \begin{pmatrix} \textbf{E} & \textbf{G}\\ \textbf{F} & \textbf{H}
\end{pmatrix} \begin{pmatrix} \textbf{A} & \textbf{0} \\
\textbf{0} & \Tilde{\textbf{A}}
\end{pmatrix}
\right]^\mathsf{T}\\
&=&\begin{pmatrix} \textbf{A}^\mathsf{T} & \textbf{0} \\
\textbf{0} & \Tilde{\textbf{A}}^\mathsf{T}
\end{pmatrix} \begin{pmatrix} \textbf{E}^\mathsf{T} & \textbf{F}^\mathsf{T}\\ \textbf{G}^\mathsf{T} & \textbf{H}^\mathsf{T}\end{pmatrix}\\
&=& \begin{pmatrix} \textbf{A}^\mathsf{T} & \textbf{0} \\
\textbf{0} & \Tilde{\textbf{A}}^\mathsf{T}
\end{pmatrix} \begin{pmatrix} \textbf{E} & \textbf{F}\\ \textbf{G} & \textbf{H}\end{pmatrix}
\end{eqnarray*}
because \textbf{E}, \textbf{F}, \textbf{G}, \textbf{H} are diagonal matrices.\\ \\
Therefore, in the Exact NL-PDHGM framework \cite{valkonen2014primal} applied to our reconstruction problem we have,
\begin{eqnarray}
    \left[\nabla \Tilde{K}_0(\bar{\textbf{x}}) \right]^* \tilde{\textbf{y}}_{0} &=& \begin{pmatrix} \textbf{A}^\mathsf{T} & \textbf{0} \\
\textbf{0} & \Tilde{\textbf{A}}^\mathsf{T}
\end{pmatrix} \begin{pmatrix} \textbf{E} & \textbf{F}\\ \textbf{G} & \textbf{H}\end{pmatrix}\begin{pmatrix} \tilde{\textbf{y}}_{0,1}\\ \tilde{\textbf{y}}_{0,2}\end{pmatrix} \nonumber\\
&=&\begin{pmatrix} \textbf{A}^\mathsf{T} \textbf{E} & \textbf{A}^\mathsf{T} \textbf{F} \\
\Tilde{\textbf{A}}^\mathsf{T}  \textbf{G} & \Tilde{\textbf{A}}^\mathsf{T}  \textbf{H}
\end{pmatrix} \begin{pmatrix} \tilde{\textbf{y}}_{0,1}\\ \tilde{\textbf{y}}_{0,2}\end{pmatrix}.
\label{eq:eq6appendix}
\end{eqnarray}
Using \eqref{eq:eq4appendix} and \eqref{eq:eq6appendix} then,
\small
\begin{eqnarray*}
    \left[\nabla \Tilde{K}_0(\bar{\textbf{x}}) \right]^* \tilde{\textbf{y}}_{0} =   \begin{pmatrix} \textbf{A}^\mathsf{T}(\underset{c=1}{\overset{2}{\sum}}(2a_{5c}\textbf{l}_w+a_{3c}\textbf{l}_i+a_{2c}\textbf{1})\odot \tilde{\textbf{y}}_{0,c})\\  \Tilde{\textbf{A}}^\mathsf{T}(\underset{c=1}{\overset{2}{\sum}}(2a_{4c}\textbf{l}_i+a_{3c}\textbf{l}_w+a_{1c}\textbf{1})\odot \tilde{\textbf{y}}_{0,c})\end{pmatrix}.
\label{eq:eq7appendix}
\end{eqnarray*}
\normalsize
Given that $\textbf{l}_w=\textbf{A}{\bar{\textbf{x}}}_w$ and $\textbf{l}_i=\Tilde{\textbf{A}}{\bar{\textbf{x}}}_{i,.}$, it leads to the solution,

\small
\begin{eqnarray*}
    \left[\nabla \Tilde{K}_0(\bar{\textbf{x}}) \right]^* \tilde{\textbf{y}}_{0} =   \begin{pmatrix} \textbf{A}^\mathsf{T}(\underset{c=1}{\overset{2}{\sum}}(2a_{5c}\textbf{A}{\bar{\textbf{x}}}_w+a_{3c}\Tilde{\textbf{A}}{\bar{\textbf{x}}}_{i,.}+a_{2c}\textbf{1})\odot \tilde{\textbf{y}}_{0,c})\\  \Tilde{\textbf{A}}^\mathsf{T}(\underset{c=1}{\overset{2}{\sum}}(2a_{4c}\Tilde{\textbf{A}}{\bar{\textbf{x}}}_{i,.}+a_{3c}\textbf{A}{\bar{\textbf{x}}}_w+a_{1c}\textbf{1})\odot \tilde{\textbf{y}}_{0,c})\end{pmatrix}
\end{eqnarray*}
\normalsize
which is equivalent to the lines \ref{grad1dyn}-\ref{grad2dyn} of the Algorithm \ref{algo1} for $\bar{\textbf{x}}=\bar{\textbf{x}}^n$ and $\tilde{\textbf{y}}_{0}=\tilde{\textbf{y}}_{0}^{n+1}$.\qquad $_\blacksquare$

\bibliographystyle{IEEEtran}
\bibliography{main}

\end{document}